# Impact of Electron-Phonon Interaction on Thermal Transport: A Review

Yujie Quan, Shengying Yue and Bolin Liao[*]

Department of Mechanical Engineering, University of California, Santa Barbara, CA 93106, USA

**Abstract**

A thorough understanding of the microscopic picture of heat conduction in solids is critical to a broad range of applications, from thermal management of microelectronics to more efficient thermoelectric materials. The transport properties of phonons, the major microscopic heat carriers in semiconductors and insulators, particularly their scattering mechanisms, have been a central theme in microscale heat conduction research. In the past two decades, significant advancements have been made in computational and experimental efforts to probe phonon-phonon, phonon-impurity, and phonon-boundary scattering channels in detail. In contrast, electron-phonon scatterings were long thought to have negligible effects on thermal transport in most materials under ambient conditions. This article reviews the recent progress in first-principles computations and experimental methods that show clear evidence for a strong impact of electron-phonon interaction on phonon transport in a wide variety of technologically relevant solid-state materials. Under thermal equilibrium conditions, electron-phonon interactions can modify the total phonon scattering rates and renormalize the phonon frequency, as determined by the imaginary part and the real part of the phonon self-energy, respectively. Under nonequilibrium transport conditions, electron-phonon interactions can affect the coupled transport of electrons and phonons in the bulk through the "phonon/electron drag" mechanism as well as the interfacial thermal transport. Based on these recent results, we evaluate the potential use of electron-phonon interactions to control thermal transport in solids. We also provide an outlook on future directions of computational and experimental developments.

---

[*] To whom correspondence should be addressed. E-mail: bliao@ucsb.edu

# I. Introduction

Heat conduction in solids plays an essential role in many technologies with significant societal impacts, such as the thermal management of electronic and optoelectronic devices [1], thermal barrier coating [2], thermoelectric materials for cooling and waste heat harvesting [3], energy storage [4], and advanced manufacturing [5]. The capability to discover and engineer solid-state materials for exceptionally high [6], low [7] or widely tunable thermal conductivities [8] is highly desirable for these applications. To this end, a complete understanding of the microscopic mechanisms of heat conduction in solids is crucial.

Electrons and phonons, or quantized lattice waves, are the major fundamental carriers of heat in solids [9]. While electrons have the dominant contribution to the thermal conductivity in metals, phonons play the primary role in semiconductors and insulators. Over the past two decades, rapid advancements in first-principles simulation and thermal metrology have led to a much more detailed understanding of the thermal transport properties of electrons and phonons. On the computational side, first-principles methods based on density functional theory (DFT), density functional perturbation theory (DFPT), and phonon Boltzmann transport equation (BTE) can now routinely compute the thermal conductivities of crystalline materials that are limited by phonon-phonon, phonon-impurity, and phonon-boundary scatterings [10]. Recent developments by Feng and Ruan have advanced the phonon-phonon scattering calculation to include the higher-order four-phonon scattering processes [11]. These calculations typically agree closely with experimental results and can resolve individual phonon modes' contribution to thermal transport. On the experimental side, recent developments of the phonon mean-free-path spectroscopy [12,13] have revealed detailed information of thermal transport contribution from phonons with different mean free paths. The combined theoretical and experimental approach has proven successful and led to the discovery of novel conduction mechanisms, such as coherent phonon transport [14] and hydrodynamic phonon transport in two-dimensional materials [15–18], as well as new materials [19–22] with unusual thermal properties for various applications.

Among all the scattering channels of phonons, the electron-phonon interaction (EPI) has been extensively studied for its impact on electrical transport properties, including electrical resistance and superconductivity. In contrast, its importance in thermal transport only started to be fully appreciated recently [23]. In the following, we will first give a historical perspective on EPI and the earlier investigations of EPI's effect on thermal transport and then review the recent computational and experimental developments in the later sections.

In principle, any states in a solid can be accurately determined by solving the Schrödinger equation involving all interactions between electrons and atomic nuclei. However, this full quantum mechanical treatment is infeasible for most condensed matter systems due to the complicated forms of interactions and a large number of involved atomic and electronic coordinates. To simplify this problem, in 1927, Born and Oppenheimer proposed that the electrons and atomic nuclei can be treated as separate quantum mechanical systems [24], the so-called Born-Oppenheimer approximation (BOA), given the fact that electrons are much lighter than atomic nuclei and that they move rapidly enough to adjust instantaneously to the much slower vibrations of the nuclei. Under BOA, the interaction term originating from the electrostatic potential generated by ionic vibrations, whose quantum description is the phonons, is dropped from the electronic Schrödinger equation. Although BOA has achieved great success in giving a reliable estimate of the total electronic energy given any atomic configurations, the dropped term, known as the EPI, is responsible for a broad range of phenomena. For example, the electrical resistance in metals at high temperatures is mainly attributed to the scattering of electrons by phonons; the attractive interaction between two electrons that form a Cooper pair, which is the origin of superconductivity, is mediated by phonons. Besides, the coupling between two electrons on the Fermi surface connected by a nesting wave vector and a phonon with a matching momentum leads to an abrupt change in the screening of lattice vibrations, which is manifested in the distortion of phonon dispersions, known as the Kohn anomaly [25].

Early efforts were focused on the influence of EPI regarding the properties of electrons in metals and semiconductors, including the explanations for the temperature dependence of electrical conductivity and electronic thermal conductivity, which have become an indispensable part of textbooks introducing transport properties of metals since the 1960s. However, less attention was paid to the lattice thermal conductivity, or the thermal conductivity mediated by phonons, in metals since it is generally believed that phonon contribution of thermal conductivity is neglectable compared with the dominance of electronic thermal conductivity. Even though thermal conduction is dominated by phonons in semiconductors, the impact of EPI on phonon transport receives less attention due to the low carrier concentration, which makes the scattering of phonons by electrons much less important than the phonon-phonon scattering. The first investigation from the phonon perspective is done by Sommerfeld and Bethe [26], who calculated the relaxation time of phonons in the presence of EPI in metals. Based on this result, Makinson gave an expression for the lattice thermal conductivity of metals with respect to temperature, where he proposed that the electrons interact equally with longitudinal and transverse phonons [27], different from Bloch's coupling scheme [28] that the electrons only interact with longitudinal phonon modes. Klemens carried out an improved numerical calculation for metals' lattice thermal conductivity at low temperatures with different strengths of EPI with transverse and longitudinal phonons [29]. As for the impact of EPI on heat conduction in semiconductors and insulators, Ziman gave an expression of the phonon lifetime due to scatterings by electrons below the Debye temperature [30]. Although these theoretical studies are based on significant simplifications (parabolic electronic bands, linear phonon dispersions, constant electron-phonon coupling matrix elements, etc.) that limit their accuracy and applicability, they significantly promoted the early understanding of the influence of EPI on phonons. These theories and the early transport experiments concluded that the impact of EPI on thermal transport in both metals and semiconductors is only significant at low temperatures when the intensity of phonon-phonon scatterings is largely suppressed [31,32].

It was not until the development of accurate first-principles calculations for phonon-phonon [33,34] and electron-phonon interactions [35] that the effect of EPI on phonon transport can be thoroughly evaluated in an extended range of materials, demonstrating that the thermal conductivity in solids can be strongly modified by the scattering of phonons by a high concentration of electrons [23], as well as the change of the phonon frequencies ("renormalization") due to EPI. The influence of EPI on phonon transport has been studied in several heavily doped materials through investigations of the dependence of their lattice thermal conductivity on carrier concentration. Furthermore, recent studies on the thermal conductivity of metals have shown that in certain transition metals, the lattice thermal conductivity is non-negligible compared with the electronic thermal conductivity [36]. In these metals, EPI plays an important role in determining phonon scattering rates and thus can impact the overall thermal transport.

It is worth noting that there have been extensive review articles written on EPI [37,38], many of which focused on the effect of EPI on electronic properties. In this review, we focus on the impact of EPI on phonon-mediated thermal transport. The paper is organized as follows: In Sec. II, we will introduce the general theory of phonon self-energy due to EPI. Sec. III will focus on the influence of EPI on phonons in thermal equilibrium. EPI can impact not only the imaginary part of the phonon self-energy, which determines the scattering rate of phonons by electrons, but also the real part of the phonon self-energy, leading to the renormalization of the phonon frequencies. Then, in Sec. IV, the impacts of EPI on the coupled transport of nonequilibrium electrons and phonons are reviewed, including the phonon and electron drag effect and the interfacial electron-phonon coupling. Finally, in Sec. V, we will give our perspectives on potential directions of future investigations. An overview of this article is given in Fig. 1.

## II. General Theory

Quantum field theory provides a systematic framework to understand the interactions between electrons and phonons, within which the many-body electron-phonon interactions are calculated by using the Green's function method. The electron Green's function $G(\mathbf{k}, \omega)$ and phonon Green's function $D(\mathbf{q}, \omega)$ can be expressed in terms of Green's functions of noninteracting electrons $G_0(\mathbf{k}, \omega)$ and phonons $D_0(\mathbf{q}, \omega)$ as well as the self-energy of electrons $\Sigma(\mathbf{k}, \omega)$ and phonons $\Pi(\mathbf{q}, \omega)$, respectively, where

$$G^{-1}(\mathbf{k}, \omega) = G_0^{-1}(\mathbf{k}, \omega) - \Sigma(\mathbf{k}, \omega), \tag{1}$$

$$D^{-1}(\mathbf{q}, \omega) = D_0^{-1}(\mathbf{q}, \omega) - \Pi(\mathbf{q}, \omega). \tag{2}$$

Based on Migdal's approximation [39], the phonon self-energy of phonon mode $\nu$ due to electron-phonon coupling can be written as

$$\Pi_\nu(\mathbf{q}, \omega_{\mathbf{q}\nu}) = \sum_{m,n} \int_{BZ} \frac{d\mathbf{k}}{\Omega_{BZ}} |g_{mn}^\nu(\mathbf{k}, \mathbf{k}+\mathbf{q})|^2 \frac{f_{m\mathbf{k}+\mathbf{q}} - f_{n\mathbf{k}}}{\varepsilon_{m\mathbf{k}+\mathbf{q}} - \varepsilon_{n\mathbf{k}} - \omega_{\mathbf{q}\nu} - i\eta}, \tag{3}$$

where the integral is over the first Brillouin zone with volume $\Omega_{BZ}$, and $\varepsilon_{n\mathbf{k}}$ and $f_{n\mathbf{k}}$ are noninteracting electronic energy and Fermi-Dirac distribution of an electron with momentum $\mathbf{k}$ and band index $n$, respectively. $g_{mn}^\nu(\mathbf{k}, \mathbf{k}+\mathbf{q}) = \sqrt{\frac{\hbar}{2m_0 \omega_{\mathbf{q}\nu}}} \left\langle \mathbf{k}n \left| \frac{\partial V}{\partial u_{\mathbf{q}\nu}} \right| \mathbf{k}+\mathbf{q}m \right\rangle$ is the matrix element of EPI involving a phonon with polarization $\nu$, wave vector $\mathbf{q}$, and the frequency $\omega_{\mathbf{q}\nu}$ as well as two electrons with band indices $m$ and $n$ and wave vectors $\mathbf{k}$ and $\mathbf{k}+\mathbf{q}$, where $\frac{\partial V}{\partial u_{\mathbf{q}\nu}}$ is the derivative of the electronic potential energy with respect to a collective ionic displacement. The imaginary part of phonon self-energy $\operatorname{Im} \Pi_\nu(\mathbf{q}, \omega_{\mathbf{q}\nu})$ is associated with the scattering rates of phonons by electrons:

$$\frac{1}{\tau_{\mathbf{q}\nu}^{ep}} = \frac{2\mathrm{Im}\Pi_\nu''(\mathbf{q}, \omega_{\mathbf{q}\nu})}{\hbar} = -\frac{2\pi}{\hbar} \sum_{mn,\mathbf{k}} |g_{mn}^\nu(\mathbf{k}, \mathbf{k}+\mathbf{q})|^2 (f_{n\mathbf{k}} - f_{m\mathbf{k}+\mathbf{q}}) \cdot \delta(\varepsilon_{n\mathbf{k}} - \varepsilon_{m\mathbf{k}+\mathbf{q}} - \omega_{\mathbf{q}\nu}), \tag{4}$$

which is directly related to the decrease of phonon thermal conductivity resulting from the increase of scattering rates of phonons by adding the electron-phonon coupling term. The $\delta$-function in Eq. (4) imposes the energy and momentum conservation conditions. On the other hand, the real part of the phonon self-energy, which gives rise to the renormalization of phonon frequencies due to EPI, can also affect phonon transport. The renormalized phonon frequency $\Omega_{\mathbf{q}\nu}$ is given by [40]

$$\Omega_{\mathbf{q}\nu}^2 = \omega_{\mathbf{q}\nu}^2 + 2\omega_{\mathbf{q}\nu}\mathrm{Re}\Pi_\nu(\mathbf{q}, \omega_{\mathbf{q}\nu}). \tag{5}$$

Researchers have developed several methods to model the potential energy $V$ in solids in order to calculate the EPI matrix elements and thus the relaxation time of phonons due to EPI. For metals, the "rigid-ion" approximation assumes that the lattice potential field surrounding each ion follows its motion rigidly [38]. Mott and Jones proposed to use the whole atomic potential of the bare ion within the Wigner-Seitz unit cell, beyond which the potential is zero [41]. Bardeen's self-consistent model [42] assumes that the perturbed potential energy of a given electron consists of not only the change of the lattice potential due to the ionic displacements but also the change caused by the spatial redistribution of the conduction electrons, which tends to screen out the perturbed potential from the distorted lattice. As for semiconductors, the seminal investigation of the phonon lifetimes (the inverse of the scattering rate) due to EPI was conducted by Ziman [30], where he used the basic model in which the phonon dispersion was depicted by the linear Debye model and the electrons were assumed to lie in a parabolic band. By further ignoring the anisotropy and limiting the applicability to highly degenerate semiconductors at low temperatures, Ziman gave an analytical expression of phonon relaxation time due to EPI:

$$\frac{1}{\tau_{ep}} = \frac{\varepsilon^2 m^{*3} v}{4\pi\hbar^4 d} \frac{kT}{\frac{1}{2}m^*v^2} \frac{\hbar\omega}{kT} - \ln\frac{1 + \exp\left[\left(\frac{1}{2}m^*v^2 - E_F\right)/kT + \hbar^2\omega^2/8m^*v^2 kT + \hbar\omega/2kT\right]}{1 + \exp\left[\left(\frac{1}{2}m^*v^2 - E_F\right)/kT + \hbar^2\omega^2/8m^*v^2 kT - \hbar\omega/2kT\right]}, \tag{6}$$

where $m^*$ is the effective mass of electrons lying in a parabolic band, $v$ is the constant phonon velocity, $d$ is mass density of the lattice, $E_F$ is the Fermi energy, and $\varepsilon$ is the EPI constant or the deformation potential, which was first proposed by Bardeen and Shockley [43]. The concept of deformation potential refers to the change of the energy of the band edges due to a static deformation of the lattice, which reflects the EPI between band edge electrons and the long-wavelength acoustic phonons. The potential change caused by lattice distortions created by long-wavelength acoustic phonons can be equivalent to the effect of a homogeneous static strain. This concept can also be extended to long-wavelength optical modes [44], which cause a relative displacement of the sublattices instead of a macroscopic deformation of the crystal. This theorem implies that the EPI information can be obtained not only from transport experiments, but also from the uniaxial stress data, and the deformation potential constant obtained from the latter significantly simplifies the calculation of the EPI matrix element.

However, the deformation potential theory ignores the contributions from phonons with shorter wavelengths and electron states not at the band edges. A more accurate way became feasible with the development of DFT-based first-principles methods in recent years. The notion of density functional theory (DFT) was proposed by Hohenberg and Kohn [45], who proved that all ground-state properties in a many-body system can be determined if the charge density distribution for the ground-state is known. Later, Kohn and Sham gave a description of the ground-state of a many-body system in terms of single-particle equations as well as an effective potential, which takes ionic potential, electron-electron interactions and many-body effects into accounts [46]. DFT methods now become indispensable in calculating electronic structures and properties, while the development of density functional perturbation theory (DFPT) paved the way for the calculations of phonon properties and electron-phonon coupling matrix elements. Within DFPT, the response of the electronic energies to infinitesimal displacements of atoms is calculated by means of the perturbation theory [47]. With the electronic and phonon properties as ingredients, however, the EPI calculations were still limited to simple systems containing few atoms in a unit cell, due to the required ultrafine sampling grid of the Brillouin zone to accurately capture the electron-phonon scattering around the Fermi surface. This requirement of high sampling grid densities is a direct consequence of the large mismatch between the electron and phonon energy scales. The recent development of an interpolation scheme based on maximally localized Wannier functions (MLWFs) [35] significantly reduced the computational cost by computing the electronic and phononic properties and the coupling matrix elements on a coarse sampling mesh, which are then interpolated to a fine mesh using generalized Fourier transforms with the MLWFs as the basis functions. This scheme tackles the challenge that the fine sampling mesh required for a converged calculation is computationally demanding. For readers interested in the technical details of the simulation methods, we refer them to relevant review articles [37,48]. In the following, we will focus our discussions on recent theoretical and experimental results regarding the effect of EPI on thermal transport in solids.

**III. Impact of EPI on Phonons in Equilibrium**

As discussed in Sec. Ⅱ, the phonon self-energy due to EPI not only contributes to the phonon scattering rate but also renormalizes the phonon frequencies. The expression of self-energy is given in Eq. 3, where the Fermi-Dirac distribution of an electron with certain momentum and band index is involved, assuming that the electrons are in their thermal equilibrium states. In fact, when calculating the EPI matrix element, the assumption that the electrons and other phonons are in equilibrium, while one phonon mode of interest deviates from equilibrium by a small amount is made. The impact of EPI with this so-called Bloch condition [49] is discussed in this section, including the reduction of the lattice thermal conductivity due to increased phonon scatterings caused by electrons as well as the renormalization of phonon dispersions and its influence on the thermal transport properties, while the EPI with nonequilibrium electrons and phonons will be discussed in the next section.

*Ⅲ.1 Imaginary Part of Phonon Self-energy: Phonon Scattering by Electrons*

The lattice thermal conductivity $\kappa_l$ can be calculated as the sum of contributions from all phonon modes: $\kappa_l = \frac{1}{3}\sum_{\mathbf{q}\nu} C_{\mathbf{q}\nu} v_{\mathbf{q}\nu}^2 \tau_{\mathbf{q}\nu}$, where $C_{\mathbf{q}\nu}$ is the heat capacity of a phonon with wave vector $\mathbf{q}$ and band index $\nu$, $v_{\mathbf{q}\nu}$ is the group velocity and $\tau_{\mathbf{q}\nu}$ is the phonon lifetime. This finite lifetime is due to phonon-phonon interactions, electron-phonon interactions, phonon-defect interactions, if any, which are combined according to Mattiessen's rule, where $\frac{1}{\tau_{\mathbf{q}\nu}} = \frac{1}{\tau_{\mathbf{q}\nu}^{\text{ph-ph}}} + \frac{1}{\tau_{\mathbf{q}\nu}^{\text{el-ph}}} + \frac{1}{\tau_{\mathbf{q}\nu}^{\text{ph-def}}}$. For a long time, the contributions to phonon scatterings by electrons were overlooked. In semiconductors with a low electron concentration, phonon scatterings by electrons are very weak. On the other hand, in metals with a high electron concentration, the lattice thermal conductivity is typically much lower than the electronic thermal conductivity.

In 2015, Liao et al. conducted systematic first-principles calculations [23] regarding EPI's effect on the lattice thermal conductivity in bulk silicon with high charge carrier concentrations up to $10^{21}$ cm$^{-3}$. This work filled the gap of understanding the role of EPI on lattice thermal conductivities of heavily-doped semiconductors and revived the research interest in investigating EPI's impact on the lattice thermal conductivity in various materials. By calculating the phonon lifetimes due to scattering with electrons (or holes) and combining them with the intrinsic lifetimes due to the anharmonic phonon-phonon interactions from first principles, they found a significant reduction of the lattice thermal conductivity at room temperature as the carrier concentration goes above $10^{19}$ cm$^{-3}$ and that the reduction can even reach up to 45% in *p*-type bulk silicon when the carrier concentration is around $10^{21}$ cm$^{-3}$, as shown in Fig. 2a. This result is surprising given the long-held belief that EPI only affects phonons strongly at temperatures much below the Debye temperature. Detailed analysis revealed that the low-frequency acoustic phonons below 5 THz are more strongly scattered by electrons than by other phonons, which explains the significant reduction of the lattice thermal conductivity since this group of phonons contribute 70% of the intrinsic thermal conductivity of silicon. Furthermore, it was pointed out that the EPI scattering rates of long-wavelength acoustic phonons in nondegenerate semiconductors scale linearly with the phonon frequency $\omega_{\mathbf{q}\nu}$, indicating that EPI is relatively more prominent for low-frequency long-wavelength phonons since the Umklapp phonon-phonon scattering rates typically scale with $\omega_{\mathbf{q}\nu}^2$. It is worth noting that despite the much stronger electron-phonon scatterings than phonon-phonon scatterings, if a semiconductor reaches a carrier concentration as high as $10^{21}$ cm$^{-3}$ via doping, typically the phonon-impurity scattering becomes the dominant phonon scattering channel. So this strong EPI effect is more relevant in materials where a high charge carrier concentration is induced without introducing atomic defects, for example by optical excitation or electrostatic gating.

In addition to silicon, other functional semiconductors also received great research interest regarding the impact of EPI on thermal transport. Fan et al. and Xu et al. both investigated the EPI effect on the lattice thermal conductivity and the thermoelectric figure of merit ZT value as a function of carrier concentration in SiGe [50,51], and reached similar conclusions that the phonon lifetimes are significantly decreased when EPI is considered and the reduction heavily depends on the carrier concentration. It is further found that the reduction of the lattice thermal conductivity of SiGe due to EPI at high carrier concentration results in a remarkable increase in the thermoelectric figure of merit ZT. Besides the typical coupling between electrons and acoustic phonons, which dominates the EPI in nonpolar crystals, the coupling between electrons and long-wavelength polar optical phonons is also of great importance in polar systems. Yang et al. investigated the influence of EPI in wurtzite GaN including the Fröhlich polar EPI for the longitudinal polar optic phonons [52] and found a 24% ~ 34% reduction for lattice thermal conductivity at 300K when the scattering of phonons by electrons is included, agreeing well with experimental results.

In parallel, there have also been emerging studies on the impact of EPI on the lattice thermal conductivity of metallic systems. Jain and McGaughey applied first-principles calculations to study mode- and temperature-dependent phonon and electron transport properties in Al, Ag, and Au by considering electron-phonon and phonon-phonon scatterings [53]. It was shown that EPI has little effect on the lattice thermal conductivity in Ag and Au, while it causes a 50% reduction in that of Al at low temperatures. Wang et al. studied lattice thermal conductivity in noble metals Cu, Ag, and Au, in d-band metals Pt and Ni, and in Al by evaluating both phonon-phonon and electron-phonon scatterings [54]. They found that EPI plays an important role in accurately determining the lattice thermal conductivity of Pt and Ni at room temperature, where the lattice thermal conductivity of Pt and Ni is reduced from 7.1 and 33.2 W/mK to 5.8 and 23.2 W/mK by considering the phonon scattering by electrons. By comparing the first-principles calculation results in those metals, they also demonstrated that the different strength of EPI scattering in different metals can be qualitatively understood as related to the electron density of states (DOS) within the Fermi window. The electronic DOS within the Fermi window determines the number of potential channels for electron-phonon scatterings and serves as a general indicator for the strength of EPI in a specific material. Tong et al. conducted comprehensive first-principles calculations in different types of metals [36] and concluded that the EPI effect on phonon thermal conductivity in transition metals and intermetallic compounds is stronger than that of noble metals, which is attributed to the larger EPI strength with a higher electron DOS within the Fermi window and higher phonon frequencies. Li and Broido et al. found up to hundred-fold reduction of the lattice thermal conductivity due to EPI in metallic transition-metal carbides NbC and TiC [55]. They attributed the significant contribution of EPI to the strong Fermi surface nesting (shown in Fig. 2c) that largely increases the available electron-phonon scattering channels, as well as a large frequency gap between acoustic and optical phonon branches that suppress the phonon-phonon scatterings. They further demonstrated that the temperature dependence of the lattice thermal conductivity deviates from the typical $1/T$ trend and becomes nearly temperature-independent when the EPI plays a dominant role (Fig. 2d). Similar properties were also found in metallic transition-metal nitrides [56] and the charge-density-wave material 1T-TaS$_2$ [57].

Besides bulk materials, the impact of EPI on phonon transport has also been studied in detail in several low-dimensional materials such as monolayer silicene and phosphorene [58], SnSe [59], Nb$_2$C [60], monolayer MoS$_2$ and PtSSe [61]. Compared with phonon scattering by electrons in three-dimensional bulk materials, this scattering mechanism in two-dimensional (2D) systems is expected to behave qualitatively differently because of different electron and phonon dispersion relations, the reduced dimensionality and the associated scattering phase space, prominent normal phonon-phonon scatterings and potential hydrodynamic phonon transport [15], and new crystal symmetries and the corresponding scattering selection rules [62]. Yue et al. adopted first-principles simulations to study the EPI effect on phonon transport in two representative 2D materials: semimetallic silicene and semiconducting phosphorene [58]. The scattering rates of acoustic phonons by electrons at low frequencies in silicene and phosphorene show different dependence on the phonon frequency compared with those in bulk materials, like silicon. In silicene and phosphorene, the scattering rates due to EPI first rise with phonon frequency below 0.3 THz and then decrease, whereas in bulk silicon the linear phonon dispersion and parabolic electronic bands give rise to the linear dependence of the phonon scattering rates on the phonon frequency for the low-frequency acoustic phonons. By developing a semi-analytical model using the deformation potential theory, they explained the impact of the reduced dimensionality and attributed the different behavior of electron-phonon scattering processes to the distinct electron and phonon dispersions in two dimensions. Furthermore, they explored the potential of controlling the lattice thermal conductivity in 2D materials via externally induced EPI by electrostatic gating, and found that over 40% reduction of the lattice thermal conductivity in 2D silicene can be achieved via inducing a charge carrier concentration around $10^{13}$ cm$^{-2}$. A similar result that the lattice thermal conductivity in monolayer SnSe decreases by as much as 30% when the carrier density exceeds $10^{13}$ cm$^{-2}$ is obtained through first-principles calculations [59]. In addition to carrier concentrations, band structures also play an important role in determining the strength of electron-phonon scatterings. It is found in $Nb_2C$ that the intensity of electron-phonon scatterings is comparable to that of phonon-phonon scattering at 300 K [60], which was attributed to the high electron DOS around the Fermi level and the special feature that the energy difference between occupied and empty electron states is on the same order of the characteristic phonon energy, making the momentum and energy conservation conditions more easily satisfied. Liu et al. also provide an example to illustrate how band structures influence the EPI process, where the effect of EPI on phonon transport in 2D $MoS_2$ and PtSSe is compared [61]. The 78% reduction of phonon lattice thermal conductivity in doped PtSSe, twice of that in equally doped $MoS_2$, is attributed to the flat band feature and thus larger electron DOS near the valence band maximum.

Experimentally, a direct verification of EPI's effect on thermal transport has been challenging. Although extensive thermal measurements of doped semiconductors have been conducted [32,63] in the past, the added phonon-impurity scattering due to doping cannot be easily separated from the EPI contribution. To directly verify the impact of EPI on thermal transport, alternative methods to induce a sufficiently high carrier concentration in a sample are needed, while in the mean time the thermal transport properties of the sample need to be evaluated. Liao et al. designed an experiment to optically generate a high concentration of electrons and holes in a thin silicon membrane, and then used the pump-probe photoacoustic spectroscopy to measure the impact of optically generated charge carriers on the lifetime of a single longitudinal acoustic phonon mode at 250 GHz [64]. In conventional pump-probe photoacoustic spectroscopy [65], the decay in the magnitude of the "echoes" generated when a phonon pulse hits the two surfaces of the sample membrane reflects the dissipation of the phonon energy due to phonon-phonon and phonon-boundary scatterings. In the modified experiment, an additional laser pulse was used to uniformly generate electron-hole pairs inside a silicon membrane, with a peak concentration up to $3 \times 10^{19}$ $cm^{-3}$. Then the additional decay of the phonon "echoes" compared to the case without photogenerated electron-hole pairs can be attributed solely to EPI. Significant additional decays of the phonon "echoes" were observed in the experiment, as shown in Fig. 3a. Converting the additional phonon decay into phonon scattering rates, the linear dependence of the EPI scattering rates on the phonon frequency was experimentally verified (Fig. 3b). More importantly, this experiment demonstrated that the scattering rate of long-wavelength acoustic phonons due to EPI exceeds that due to phonon-phonon scatterings when the carrier concentration is above $10^{19}$ $cm^{-3}$. Although this experiment provided direct evidence of strong EPI impact on phonon transport, it can only probe the response of one individual phonon mode. Zhou et al. generalized this result by directly measuring the impact of EPI on the thermal conductivity of a silicon membrane at room temperature by combining the additional photo-excitation of electron-hole pairs and the transient thermal grating (TTG) technique [66]. TTG is a well established method to characterize the in-plane thermal transport of membranes [67]. By incorporating an additional optical pulse to generate a high concentration of charge carriers in the membrane, the impact of EPI on the lattice thermal conductivity of the membrane can be directly evaluated using TTG. A large change of the thermal decay time in the TTG measurement was observed when extra electron-hole pairs were generated in the sample, as shown in Fig. 3c. A suppression of the lattice thermal conductivity by as much as 30% was observed when the photo-generated carrier concentration reached $7.8 \times 10^{19}$ cm$^{-3}$, in quantitative agreement with theory (Fig. 3d). This experiment directly demonstrated the non-negligible impact of EPI on the lattice thermal conductivity of semiconductors.

### *III.2 Real Part of the Phonon Self-energy: Phonon Frequency Renormalization*

The real part of the phonon self-energy due to EPI modifies phonon frequencies obtained from the adiabatic Born-Oppenheimer approximation, and this renormalized phonon frequency is given by Eq. (5). Since the phonon frequencies affect the phonon group velocity, the phonon heat capacity, and the phonon scattering phase space, the phonon frequency renormalization due to EPI can potentially affect the lattice thermal conductivity, particularly if significant phonon softening is induced. Phonon softening due to resonant bonding [68] and ferroelectric instability [69] are well known to cause abnormally low lattice thermal conductivity due to enhanced phonon-phonon scattering phase space. Specific to EPI, strong phonon softening can be induced by the so-called Kohn anomaly [25]. Kohn showed in 1957 that when two electronic states with wave vectors $\mathbf{k_1}$ and $\mathbf{k_2}$ on the Fermi surface are parallelly nested by a phonon with a wave vector $\mathbf{q} = \mathbf{k_2} - \mathbf{k_1}$, a discontinuity in the derivative of the phonon dispersion at this wave vector will arise. In practice, this discontinuity usually results in an abrupt "dip" in the phonon dispersion at the nesting wave vector [70]. Physically, the Kohn anomalies are produced by the abrupt change in the screening of lattice vibrations with a specific wavelength by conduction electrons due to a singular EPI when the participating electrons and phonons satisfy the nesting condition. Strong Kohn anomalies can lead to structural instability and charge-density-wave (CDW) transitions [71] when the "dip" in the phonon dispersion reaches zero frequency.

In general, Kohn anomalies are more prominent in 1D and 2D conductors, given the stronger divergences of the dielectric screening function at the nesting wavevector [71]. For this reason, Kohn anomalies have been extensively studied in low dimensional and layered zero-gap materials, such as graphene [72,73], graphite [74], and surface states of topological insulators [75]. An intriguing case emerges in the recently discovered 3D topological Dirac and Weyl semimetals, which feature 3D linear dispersions at discrete points in the Brillouin zone (Dirac or Weyl points). Since the Fermi surface of these 3D materials simply consists of isolated Dirac or Weyl points in the Brillouin zone, unusual Kohn anomalies associated with these Dirac/Weyl points were predicted to exist. Nguyen et al. reported the chiral Kohn anomaly in a 3D Weyl semimetal

TaP [76]. They demonstrated the existence of Kohn anomalies in TaP by field-theoretical calculations and inelastic x-ray and neutron scattering experiments, where phonon softening was observed at wavevectors matching the distance between Weyl points. As for Dirac semimetals, Yue et al. reported the existence of ultrasoft optical phonons in topological Dirac semimetal $Cd_3As_2$ by first-principles lattice dynamics calculations and temperature-dependent high-resolution Raman spectroscopy [77]. The soft optical phonons are attributed to potentially strong Kohn anomalies associated with the Dirac nodes, as shown in Fig. 4a. Phonon softening at two locations in the Brillouin zone were observed: one near the zone center, and the other near the wave vector $\mathbf{q}_0$ that matches the distance between two Dirac nodes in the Brillouin zone. Phonon softening occurring at these two locations is a strong evidence that they are caused by Kohn anomalies: one associated with intranode electron-phonon scattering, and the other associated with internode electron-phonon scattering [76]. Moreover, the phonon softening was shown to be highly sensitive to the electronic smearing parameter $\sigma$, which reflects the broadening of the electronic Fermi-Dirac distribution and effectively acts as a temperature of the electrons. In addition, it is found that the ultralow lattice thermal conductivity of $Cd_3As_2$, in the range of 0.3 to 0.7 W/mK at 300 K [78], is due to the increased scattering phase space of heat-carrying acoustic phonons induced by soft phonon modes. The highly temperature dependent soft phonon frequency also explains the unusual temperature dependence of the lattice thermal conductivity as shown in Fig. 4b. Along this line, Yue et al. further showed that these soft phonons can be sensitively tuned across topological phase transitions driven by external stimuli, where the semimetallic phase appear at the phase transition point [79]. For example, $ZrTe_5$ undergoes a topological phase transition when subjected to a hydrostatic pressure of 5 GPa, where the electronic band gap closes and a semimetallic phase emerges, as shown in Fig. 4c. Associated with the emergence of the semimetallic phase, significant phonon softening at the zone center and near the Y point appears, as shown in Fig. 4d, signaling strong Kohn anomalies. The same phonon softening effect was also found near a chemical-composition-induced topological phase transition in $Hg_{1-x}Cd_xTe$ [79]. These emergent Kohn anomalies associated with the closing of the band gap can be potentially utilized as a means to realize sensitive switching of the thermal conductivity via topological phase transitions driven by external stimuli.

**IV. Impact of EPI on Phonons under Nonequilibrium Conditions**

So far, we have focused our discussion on EPI when electrons and phonons are in thermal equilibrium, described by Fermi-Dirac and Bose-Einstein statistics, respectively. In real transport scenarios, however, electrons and phonons will generally deviate from their equilibrium states. In materials hosting both phonons and conducting electrons, the transport processes of electrons and phonons driven by either a temperature gradient or an electric field are closely coupled. A classic example of coupled electron-phonon transport is the "phonon drag" effect [80], where a phonon heat flow driven by a temperature gradient can "drag" an electron flow via EPI, leading to an enhanced Seebeck effect. Reversely, an electron flow driven either by temperature or electric field can also "drag" a phonon flow, leading to coupled transport ("electron drag"). A precise description of the coupled electron-phonon transport requires the solution of the coupled electron-phonon Boltzmann transport equations:

$$\mathbf{v}_\alpha(\mathbf{k}) \cdot \nabla_\mathbf{r} f_\alpha(\mathbf{k}) + \frac{\mathbf{F}}{\hbar} \cdot \nabla_\mathbf{k} f_\alpha(\mathbf{k}) = \left(\frac{\partial f_\alpha(\mathbf{k})}{\partial t}\right)_{\text{e-ph}}, \tag{7}$$

$$\mathbf{v}_\lambda(\mathbf{q}) \cdot \nabla_\mathbf{r} n_\lambda(\mathbf{q}) = \left(\frac{\partial n_\lambda(\mathbf{q})}{\partial t}\right)_{\text{ph-ph}} + \left(\frac{\partial n_\lambda(\mathbf{q})}{\partial t}\right)_{\text{e-ph}}. \tag{8}$$

where $\mathbf{v}_\alpha(\mathbf{k})$ and $\mathbf{v}_\lambda(\mathbf{q})$ are group velocities of electrons and phonons with wave vectors $\mathbf{k}$ and $\mathbf{q}$, and band indices $\alpha$ and $\lambda$, $f_\alpha$ and $n_\lambda$ are the nonequilibrium distribution functions of electrons and phonons, respectively. The scattering terms on the right-hand side of Eqs. (7) and (8) include electron-phonon and phonon-phonon scattering processes that couple the transport of electrons and phonons. When electrons and phonons only deviate slightly from their thermal equilibrium distribution, i.e. under near-equilibrium transport conditions, Eqs. (7) and (8) can be expanded in a perturbative series in terms of the small deviations of $f_\alpha$ and $n_\lambda$ from their equilibrium values [48]. With this approach, Zhou et al. calculated the phonon drag contribution to the Seebeck coefficient of silicon using first principles methods [81], as shown in Fig. 5a. When the electron-phonon scattering is much weaker than phonon-phonon scattering, e.g. in 2D graphene, electrons and phonons tend to equilibrate among themselves with different temperatures. In this scenario, the coupled transport of electrons and phonons can be approximately treated using two-temperature or multi-temperature models [82,83], such as the case shown in Fig. 5b. This nonequilibrium effect between electrons and phonons (and even among different phonon branches) is particularly important when interpreting thermal measurements using optical probes [83], such as Raman spectroscopy and TTG.

Recently, Protik and Broido developed a numerical approach to directly solve the coupled electron-phonon Boltzmann transport equations, Eqs. (7) and (8), with phonon-phonon scattering matrix elements calculated using first-principles methods and electron-phonon scattering matrix elements calculated from analytical models [84]. With this method, they were able to accurately compute the phonon drag Seebeck coefficient of n-type GaAs. They also studied the mutual drag effect on the electron mobility and the lattice thermal conductivity in n-type GaAs and concluded that the electron mobility is significantly enhanced by phonon drag while the lattice thermal conductivity is only weakly affected by electron drag. In a follow-up work, Protik and Kozinsky further developed this method with both phonon-phonon and electron-phonon scattering matrix elements calculated from first principles [85]. With this improved approach, they found that the room-temperature lattice thermal conductivity in n-doped cubic SiC is increased by 8% due to electron drag with a high carrier concentration of $10^{20}$ cm$^{-3}$, which, however, is fully suppressed when impurity scattering is included (Fig. 5c). This is due to the fact that only acoustic phonons with very long wavelengths and polar optical phonons are strongly affected by the electron drag, while the lattice thermal conductivity has contributions from a much wider phonon spectrum (Fig. 5d). It remains an open question whether there are materials whose lattice thermal conductivity can be significantly enhanced by electron drag.

Another important scenario where thermal transport is affected by nonequilibrium electron-phonon coupling is interfacial thermal transport between insulators and metals (or doped semiconductors). Majumdar and Reddy first theoretically studied the contribution of EPI to the thermal conductance of metal-nonmetal interfaces and showed the contribution can be significant at high temperatures [86]. Combining Boltzmann transport equations and the two-temperature model for electrons and phonons, Wang et al. demonstrated that the electron-phonon nonequilibrium near metal-dielectric interfaces largely adds to the total interfacial thermal resistance and can be reduced by inserting an interlayer with relatively strong electron-phonon coupling [87]. Currently, there is no fully first-principles treatment of coupled electron-phonon transport across interfaces.

**V. Summary and Outlook**

In summary, we reviewed the recent progress in understanding the impact of EPI on thermal transport in solids. Under thermal equilibrium conditions, EPI can lead to additional phonon scattering and renormalize the phonon frequency, both of which can strongly influence the lattice thermal conductivity. Under nonequilibrium conditions, the coupled transport of electrons and phonons can significantly affect the thermal and thermoelectric transport properties, including the Seebeck coefficient, electron mobility and lattice thermal conductivity.

Further developments are required to gain more detailed understanding of the EPI effect on thermal transport, as well as explore novel applications based on EPI such as solid-state thermal switches. Experimentally, inelastic x-ray and neutron scattering (IXS and INS) techniques now have the energy and momentum resolution to map the frequency and linewidth (directly related to the phonon lifetime) of each individual phonon modes in solids [69,76]. To examine the impact of EPI on the phonon linewidth, however, IXS and INS experiments with additional optical excitation or electrostatic gating need to be developed. Similarly, to directly evaluate the impact of EPI on the steady-state thermal transport processes, steady-state thermal measurement techniques [88] need to be combined with optical excitation or electrostatic gating to adjust the charge carrier concentration of the sample without introducing defects. New spectroscopic methods, e.g. based on electron energy loss spectroscopy [89], ultrafast electron diffraction [90] or ultrafast electron microscopy [91], are highly desirable to directly probe the electron-phonon nonequilibrium near surfaces and interfaces. Theoretically, more systematic studies are needed to identify the key criteria beyond the electron DOS to select materials with particularly strong or weak EPI effect on thermal transport. New first-principles methods are needed to fully understand the coupled electron-phonon transport across interfaces and in nanostructures where the classical size effect becomes appreciable. Furthermore, the nonequilibrium of electrons and phonons also affects the scattering rate calculations, which require the electron and phonon distribution functions. It is experimentally known that electron-phonon interaction is weakened in strongly optically pumped semiconductors [92], when the high density of photo-excited electrons block certain electron-phonon scattering channels. Theoretical development along this direction is particularly relevant to the performance of optoelectronic and photovoltaic devices. For practical applications, the EPI effect on thermal transport can be potentially utilized to develop solid-state thermal switches, whose thermal conductivity can be sensitively controlled by external stimuli. Since electrons can effectively and swiftly respond to external electrical and magnetic fields, indirect control of the electron properties can potentially modulate the thermal conductivity through EPI [58]. For this purpose, materials with particularly strong EPI effect on thermal transport still need to be identified, which can only be achieved by continued combined advancements of theoretical and experimental methods.

**Acknowledgments**

This work is based on research supported by the National Science Foundation under the award number CBET-1846927. Y. Q. acknowledges the support from the Graduate Traineeship Program of the UCSB NSF Quantum Foundry via the Q-AMASE-i program under award number DMR-1906325.

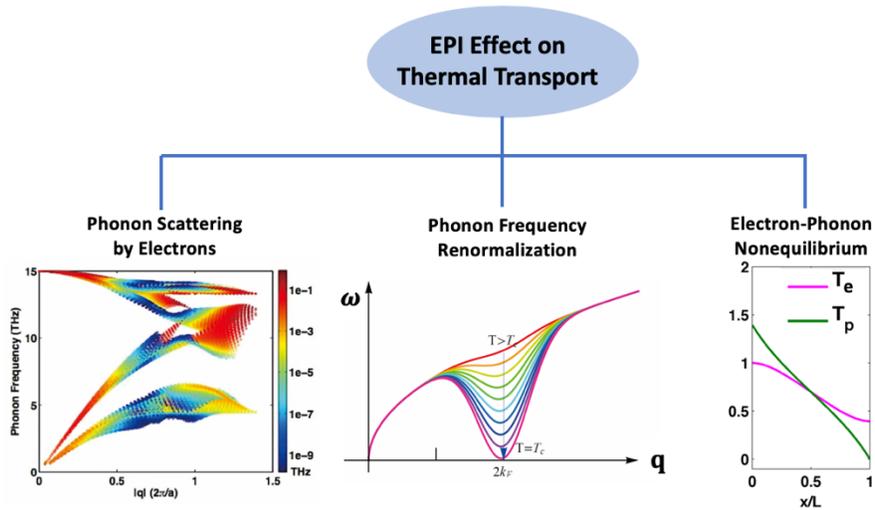

**Figure 1. Overview of the EPI effect on thermal transport.** EPI can lead to additional phonon scatterings by electrons (left panel) and renormalize the phonon frequencies (center panel), both of which can affect thermal transport. Furthermore, coupled transport of nonequilibrium electrons and phonons (right panel) can significantly alter thermal and thermoelectric transport properties. Left panel is adapted from Ref. [23] with permission. Copyright: American Physical Society.

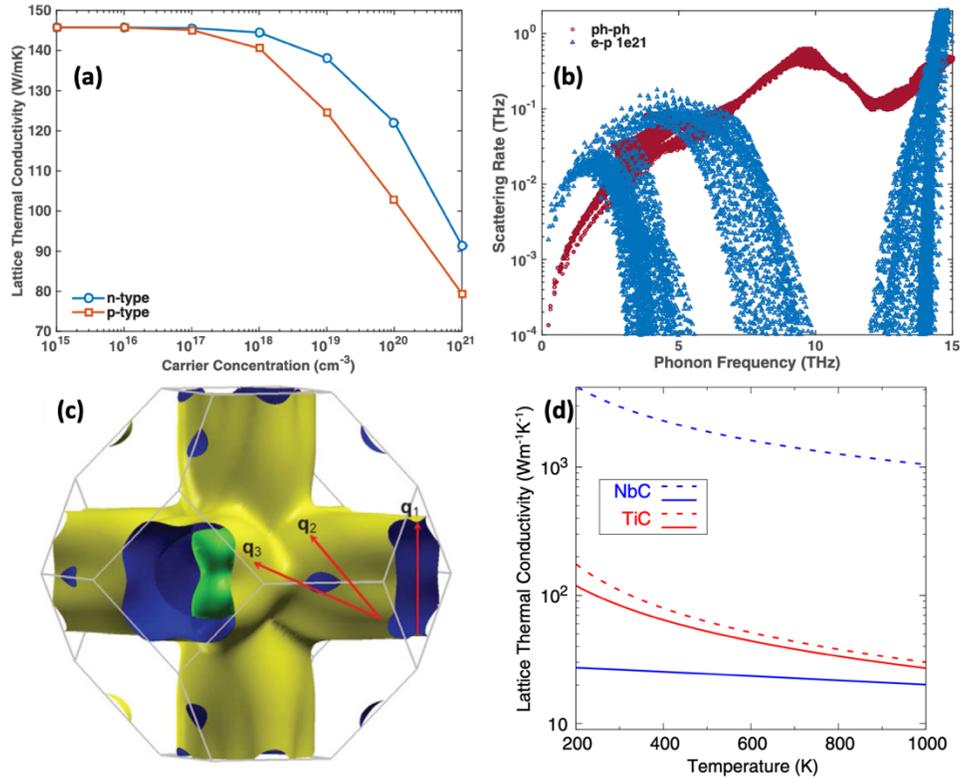

**Figure 2. First-principles simulation of phonon-electron scattering rates.** (a) Significant reduction of the lattice thermal conductivity of silicon due to EPI as a function of the charge carrier concentration. (b) Comparison of phonon scattering rates due to intrinsic phonon-phonon scattering (red circles) and electron-phonon scattering (blue triangles) with a charge carrier concentration of $10^{21}$ cm$^{-3}$ in p-type silicon. (a) and (b) are adapted from Ref. [23] with permission. Copyright: American Physical Society. (c) The Fermi surface of the transition-metal carbide NbC, showing significant nesting that enhances the possible EPI channels. (d) The temperature dependence of the lattice thermal conductivity of NbC and TiC (dashed lines: only include phonon-phonon scattering and phonon-isotope scattering; solid lines: include phonon-phonon scattering, phonon-isotope scattering and phonon-electron scattering). Strong EPI in NbC leads to a nearly temperature-independent lattice thermal conductivity. (c) and (d) are adapted from Ref. [55] with permission. Copyright: American Physical Society.

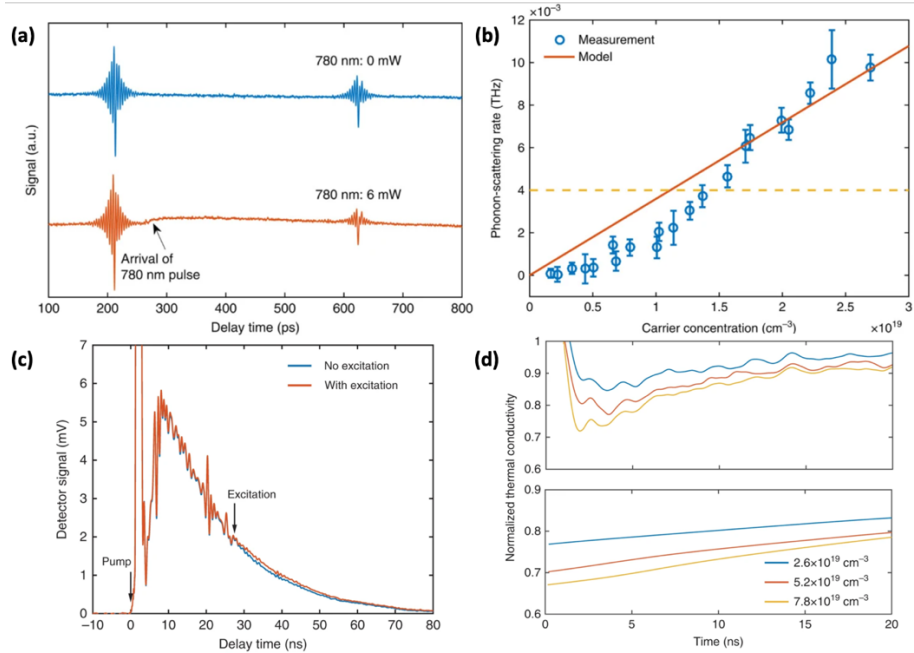

**Figure 3. Experimental measurements of the phonon scattering rates due to EPI.** (a) The phonon "echoes" in a pump-probe photoacoustic spectroscopic experiment showing the significant impact of an additional photo-excitation pulse on the magnitude of the phonon echo. This additional decay is due to EPI caused by photo-excited charge carriers. (b) The measured scattering rate of a 250 GHz longitudinal acoustic phonon due to EPI as a function of photo-generated charge carrier concentration, and its comparison to a theoretical model. (a) and (b) are adapted from Ref. [64] with permission. Copyright: Nature Publishing Group. (c) The impact of the photo-excited electron-hole pairs on the thermal decays measured in a TTG experiment. (d) The transient lattice thermal conductivity (normalized to the intrinsic value) extracted from the TTG experiment with different photo-excited charge carrier concentrations (upper panel), in comparison to a theoretical model (lower panel). (c) and (d) are adapted from Ref. [66]. Copyright: Nature Publishing Group.

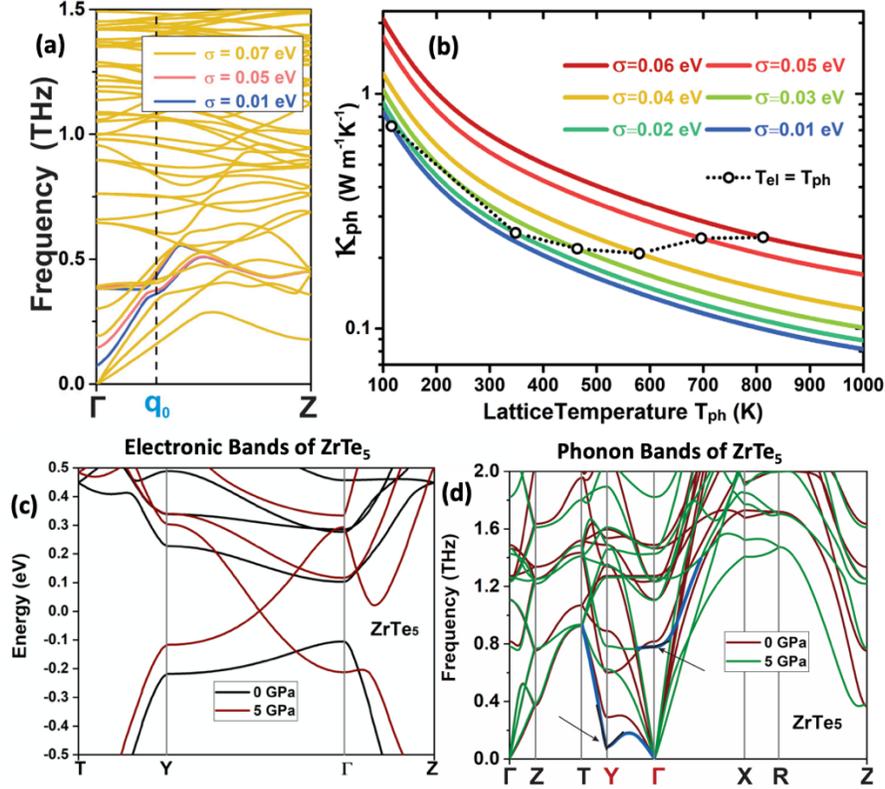

**Figure 4. Phonon softening due to Kohn anomalies in topological semimetals.** (a) Phonon dispersion of topological Dirac semimetal $Cd_3As_2$. $\sigma$ is the electronic smearing parameter. (b) The calculated lattice thermal conductivity of $Cd_3As_2$ with different lattice temperatures and electronic smearing parameters. (a) and (b) are adapted from Ref. [77] with permission. Copyright: American Physical Society. (c) The electronic band structure of $ZrTe_5$ showing a pressure-driven topological phase transition at 5 GPa. (d) The phonon softening in $ZrTe_5$ associated with the electronic topological phase transition. (c) and (d) are adapted from Ref. [79] with permission. Copyright: American Physical Society.

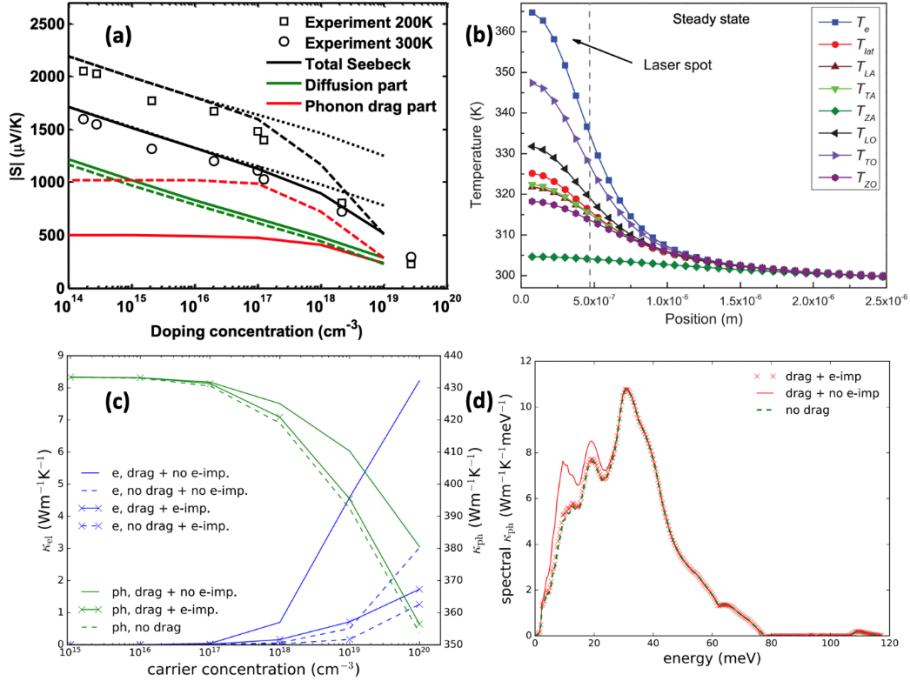

**Figure 5. The impact of EPI on nonequilibrium coupled electron-phonon transport.** (a) The calculated phonon drag contribution to the Seebeck coefficient of silicon compared to experimental results. Adapted from Ref. [81] with permission. Copyright: U.S. National Academy of Sciences. (b) The nonequilibrium among electrons and different phonon branches with different temperatures in graphene under optical illumination. Adapted from Ref. [82] with permission. Copyright: American Physical Society. (c) The calculated electronic and lattice thermal conductivity in cubic SiC with and without the mutual electron-phonon drag effect and impurity scatterings. (d) The impact of electron drag on the spectral thermal conductivity of phonons with different frequencies (energies) in cubic SiC. (c) and (d) are adapted from Ref. [85] with permission. Copyright: American Physical Society.